# Anomalous spectral evolution with bulk sensitivity in BiPd

Arindam Pramanik[1], Ram Prakash Pandeya[1], Khadiza Ali[1], Paolo Moras[2], Polina M. Sheverdyaeva[2], Carlo Carbone[2], Bhanu Joshi[1], A. Thamizhavel[1], S. Ramakrishnan[1] and Kalobaran Maiti[1,a)]

[1]Department of Condensed Matter Physics and Materials Science, Tata Institute of Fundamental Research, Homi Bhabha Road, Colaba, Mumbai, 400005, India.
[2]Istituto di Struttura della Materia, Consiglio Nazionale delle Ricerche, Area Science Park, I-34012 Trieste, Italy

a) Corresponding author: kbmaiti@tifr.res.in

**Abstract.** We investigate the electronic structure of a noncentrosymmetric superconductor, BiPd using photoemission spectroscopy with multiple photon energies ranging from ultraviolet to hard *x*-ray. Experimental data exhibit interesting difference in the surface and bulk electronic structures of this system. While the surface Bi core level peaks appear at lower binding energies, the surface valence band features are found at the higher binding energy side of the bulk valence band; valence band is primarily constituted by the Pd 4*d* states. These changes in the electronic structure cannot be explained by the change in ionicity of the constituent elements via charge transfer. Analysis of the experimental data indicates that the Bi-Pd hybridization physics plays the key role in deriving the anomalous spectral evolution and the electronic properties of this system.

## INTRODUCTION

The compound, BiPd has attracted tremendous attention of scientific community because of the finding of superconductivity in a system with lack of inversion symmetry in its crystal structure, which enables it to have spin-singlet and spin-triplet mixing of Cooper pairs [1]. Additionally, due to the expectation of the Dirac surface states in the surface electronic structure, it is being considered to be a potential candidate for topological superconductor. Recent angle resolved [2-4] and spin resolved [2] photoemission studies have provided evidence of the Dirac like surface states on the cleaved surface of single crystalline BiPd. The scanning tunneling spectroscopy [5] and quantum oscillation [6] measurements also suggest topologically nontrivial nature of the material. BiPd forms in orthorhombic crystal structure at high temperature with the space group $Cmc2_1$, known as $\beta$-BiPd. Below 210 °C, it assumes monoclinic structure with the space group $P2_1$, known as $\alpha$-BiPd, which lacks inversion symmetry and the unit cell contains four inequivalent Bi and Pd atoms [1]. Here, we studied the evolution of the electronic structure employing high resolution photoemission spectroscopy with increasing bulk sensitivity of the technique. In addition, core level spectra were collected at different temperatures to investigate the temperature evolution of the electronic structure.

## EXPERIMENT

High quality single crystals of BiPd were grown using modified Bridgman technique. Hard x-ray photoemission (HAXPES) studies were carried out at P09 beamline, Petra III, Hamburg, Germany. The energy resolution of 200 meV could be achieved for 5946.6 eV photon energy and hence, we used this energy for HAXPES studies. Studies using ultraviolet photons were done at VUV photoemission beamline at Elettra, Trieste, Italy with energy resolution set at 15 meV. For the conventional *x*-ray photoemission spectroscopy (CXPS) studies, we used an Al $K\alpha$ laboratory source from Specs GmbH, Germany and the instrumental resolution was set to 450 meV for optimum spectral intensity. Prior to the photoemission measurements, the samples were oriented using Laue diffraction method and cleaved *in-situ* along the *b* axis, which is the preferred cleaving direction of the crystals, to expose a clean surface for the photoemission measurements. To increase the surface sensitivity in the HAXPES measurements (highly bulk sensitive with mean escape depth ~ 40 Å), electron emission angle with respect to the sample surface normal was varied [7]. At higher emission angle, the photoelectrons need to travel longer path inside the sample before ejecting through the sample surface to reach the detector, which reduces the signal intensity from the bulk and thereby, makes the technique relatively more surface sensitive.

## RESULTS AND DISCUSSIONS

In Fig. 1(a), we show the Bi 4*d* core level HAXPES spectrum collected at normal emission geometry at 50 K exhibiting two intense features at 440.4 eV and 464.1 eV binding energies. These are identified as the spin-orbit split $4d_{5/2}$ and $4d_{3/2}$ core level peak. In order to identify the other contributions appearing at higher binding energies, we have fit the experimental spectrum using asymmetric peaks of Doniach-Šunjić lineshape; the simulated spectrum (red line superimposed over the experimental data represented by open circles) shows a good agreement with the experiment. Core level spectra often exhibit multiple features due to multiplet effects [8] and/or charge transfer effects [9]. Here, one can capture the spectral lineshape of each of the features with one asymmetric peak. The asymmetry can be explained as follows. Along with the photoexcitation of the core electrons due to the perturbation by the incident photon beam, valence electrons can also get excited across the Fermi level leading to a reduction in the kinetic energy of the core photoelectrons. Corresponding contributions will appear at the higher binding energy side of the peak, which is the asymmetry in the lineshape. Thus, presence of such lineshape indicates that there is finite density of states at the Fermi level that allows such excitations. From the fitting we find that full width at half maximum (FWHM) of the $4d_{3/2}$ peak (~ 4.56 eV) is larger than $4d_{5/2}$ peak (~ 4.32 eV); slightly higher FWHM for $4d_{3/2}$ is expected as the higher binding energy of $4d_{3/2}$ core level relative to $4d_{5/2}$ core level energy allows an additional decay channel of the $4d_{3/2}$ core hole involving $N_4N_5V$ Auger transition giving rise to a relatively larger lifetime broadening.

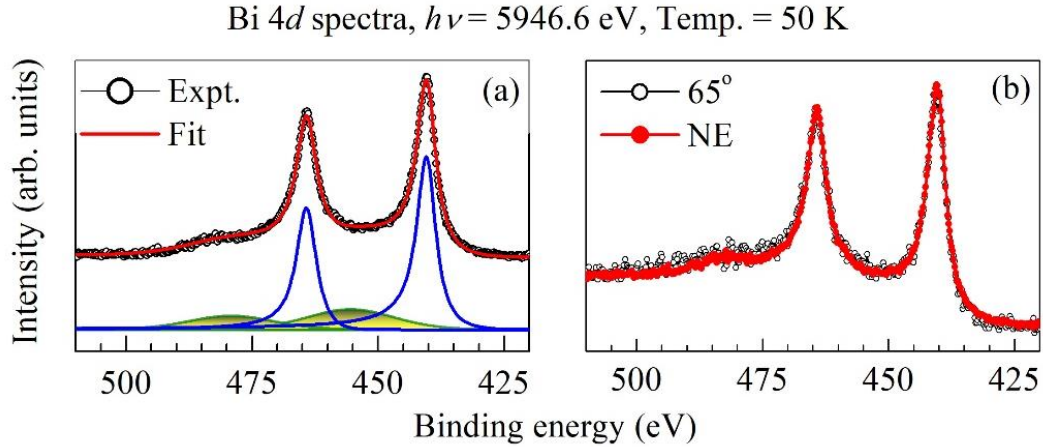

**Figure 1.** (a) Bi 4*d* core level spectra. Upper panel shows the simulated spectrum (red line) overlain onto the experimental data (open black circle). Four component peaks are indicated in the lower panel. Two lower intensity peaks are identified as loss features due to plasmon excitations. (b) Both normal and 65° angled emission data are plotted together exhibiting identical lineshape and intensities.

The two broad peaks with low intensity (see the shaded area plot in Fig. 1(a)) are identified as loss features arising due to the excitation of the collective modes along with the photoexcitation of $4d_{5/2}$ and $4d_{3/2}$ core electrons, respectively. They are situated about 15 eV away from their respective main peaks. Confirmation of these peaks as loss features comes from the fact that they appear in all the core levels studied (not shown here for brevity). Energy loss of this amount could be due to plasmon excitations; the collective excitation of Fermi sea [7].

In Fig. 1(b), we show both 65° angled and normal emission spectra superimposed on each other. Excellent degree of overlap indicates that either there is no difference between the surface and bulk electronic structures or the angled emission geometry does not possess enough surface sensitivity to reveal the differences. Additionally, it also suggests that if the surface electronic structure is different, such effect is confined within the top few layers near the surface layer.

In order to probe the effect further, we used ultraviolet photons, where the surface sensitivity is much higher [10]. Bi 5$d$ core level spectrum shown in Fig. 2 (lower panel) was collected with 45 eV photon energy at 15 K. Each spin-orbit split peak is clearly seen to be composed of two distinct features. Comparison with the hard *x*-ray spectrum (Fig.2, upper panel) shows that the higher binding energy component (red line) is arising from the core levels of bulk atoms. The other peak (green line) appearing at lower binding energy is identified as the surface feature. The energy difference between the surface and bulk peak positions is about 0.3 eV. This indicates that the surface and bulk electronic structure in this system is indeed very different although it was not apparent in the HAXPES data at angled emission.

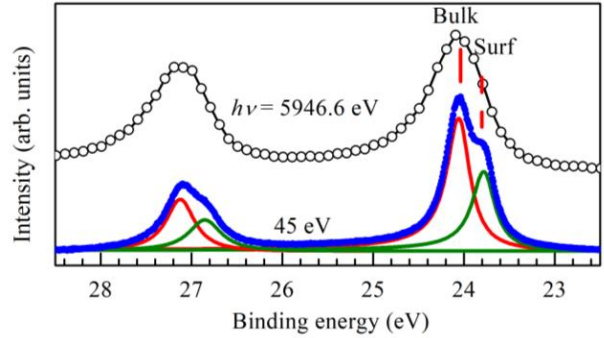

**FIGURE 2.** Bi 5$d$ spectrum. Upper panel shows the hard x-ray spectrum taken at 50 K. Lower panel displays the same spectrum collected with 45 eV photons at 15 K. Each spin-orbit split peak is seen to be composed of two peaks.

In Fig. 3, we show the valence band spectra collected with Al $K\alpha$ (1486.6 eV) source at 195 K – it is to note here that the change in temperature does not have much influence in the core level spectra as well as the valence band spectra studied here. Different angled emission geometries have been used to delineate the bulk and surface contributions in the valence band. In Fig. 3(a), we show normal, 45° and 70° emission spectra. With the increase in emission angle, the surface sensitivity of the technique increases, and we observe significant increase in spectral intensity at the higher binding energy side. To demonstrate the change in intensity with better clarity, we show the difference of the spectra with respect to the normal emission spectrum in Fig. 3(b) (angled emission – normal emission). The difference spectra exhibit double peak structure as also seen in the normal emission spectra indicating a shift of the surface valence band towards higher binding energies.

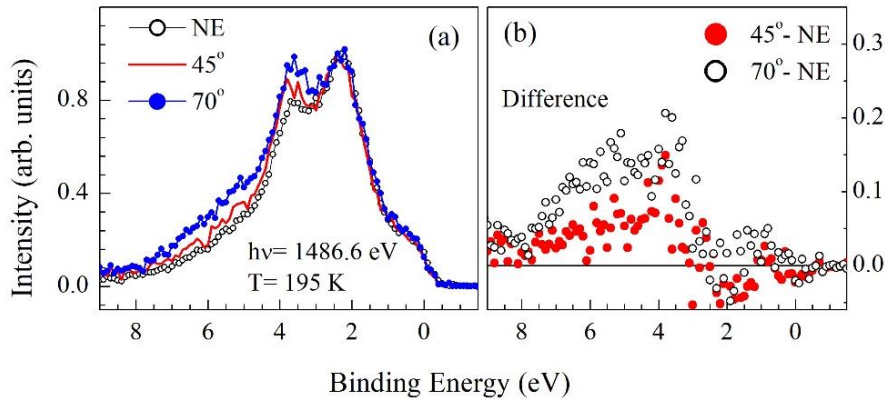

**FIGURE 3.** Valence band spectra of BiPd. (a) Spectrum collected at normal, 45° and 70° emission geometries using Al $K\alpha$. (b) Subtracted spectra exhibiting increase in intensity at higher binding energy for the angled emissions.

Such spectral shift can occur due to enhancement of the binding energy of the valence electrons [11]. From the core level spectra, we found that the surface core level appears at lower binding energies. Thus, a rigid shift of the Fermi level would shift the valence band to lower energies. However, we observe opposite energy shift which cannot be captured considering charge transfer as enhancement of Pd 4$d$ population would make Pd more negative resulting to a shift in opposite direction.

Cleaved surface is expected to have less number of Bi atoms as the Bi atoms will get distributed to both the surfaces created due to cleaving. Thus, the hybridization between remaining surface Bi atoms will be more attracted towards Pd sites leading to an enhancement of Bi-Pd hybridization. We believe that the shift of the surface valence band towards higher binding energy is the reflection of enhanced hybridization at the surface relative to the bulk leading to a shift of the bonding states towards higher energies.

To probe the temperature evolution of the electronic structure, Bi $5d$ core level spectra were measured at multiple temperatures. Since the Debye temperature, $\theta_D$ of BiPd is 169 K [1], which is accessible for our experiments, measurements were carried out well below and above $\theta_D$. In Fig. 4, we show the Bi $5d$ spectra at three different temperatures. With the increase in temperature, the width of the bands are expected to increase due to higher rate of electron-phonon scattering and phonon excitations. But we found no discernible change in the width of the Bi $5d$ core level spectra over a wide temperature range. Since $\theta_D$ is only 169 K, which gives 14.4 meV as the highest phonon energy in the system, we think the instrumental resolution (200 meV) masks the thermal effects on the sample.

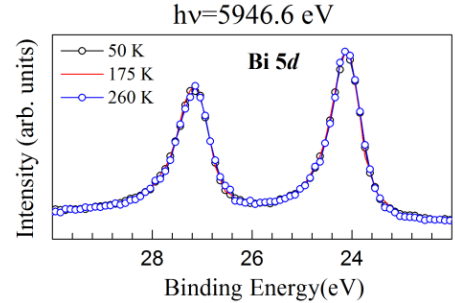

**Figure 4.** Bi $5d$ spectrum taken at various temperatures.

## CONCLUSIONS

In summary, we have studied the electronic structure of high quality single crystalline BiPd using both $x$-ray and ultraviolet photoemission spectroscopy. Experimental spectra exhibit distinct peaks corresponding to the surface and bulk electronic structures. Bi $5d$ Surface features are found to be shifted towards lower binding energy by 0.3 eV. In contrast, surface valence band (valence bands are constituted primarily of Pd $4d$ states) shows shift towards higher binding energy. Such anomalous energy shift can be attributed to higher degree of Bi-Pd hybridization at the surface. Temperature is found to play negligible effect on the photoemission spectra.

## ACKNOWLEDGMENTS

The authors acknowledge financial support from the DST-DESY project to perform the experiments at P09 beamline at PETRA III, Hamburg, Germany and Dr. Indranil Sarkar for his help during the measurements. Financial support from the Department of Atomic energy, Government of India for the experiments at Elettra is thankfully acknowledged. KM acknowledges financial assistance from the Department of Science and Technology, government of India under the J. C. Bose Fellowship program and the Department of Atomic Energy under the DAE-SRC-OI Award program.